\documentclass[namedreferences]{solarphysics}

\usepackage[optionalrh]{spr-sola-addons} 
\usepackage{graphicx}        
\usepackage{color}           

\begin{document}

\begin{article}
\tracingmacros=2
\begin{opening}
 \title{New Observations of Balmer Continuum Flux in Solar Flares\\
   {\textrm Instrument Description and First Results}}

\author[addressref={aff1},email={pkotrc@asu.cas.cz}]{P.~Kotr\v{c}}\sep
\author[addressref={aff1,aff2},corref,email={prochazkaondrej@seznam.cz}]{O.~Proch\'{a}zka}\sep
\author[addressref=aff1,email={pheinzel@asu.cas.cz}]{P.~Heinzel}\sep

\address[id=aff1]{Astronomical Institute, Academy of Sciences of the Czech Republic,
             Fri\v{c}ova 298, 25165 Ond\v {r}ejov, Czech Republic}
\address[id=aff2]{Faculty of Mathematics and Physics,  Charles University,
            V Hole\v{s}ovi\v{c}k\'{a}ch 2, 180 00 Praha 8, Czech Republic}

\runningauthor{P. Kotr\v{c} \textit{et al.}} \runningtitle{Balmer Continuum
Flux in Solar Flares}


  \begin{abstract}
Increase in the
Balmer continuum radiation during solar flares was predicted by various authors
but never firmly confirmed observationally using ground-based slit spectrographs. Here we describe
a new post-focal instrument \-- Image Selector \-- enabling to measure the Balmer
continuum flux from the whole flare area, in analogy of successful detections of
flaring dMe stars. The system was developed and put into
operation at the horizontal solar telescope HSFA-2 of the Ond\v{r}ejov Observatory.
We measure the total flux by a fast spectrometer from a limited but well defined region on the solar disk.
Using a system of diaphragms,
the disturbing  contribution  of a bright solar disk can be eliminated as much as
possible. Light curves of the measured flux in the spectral
range 350 - 440 nm are processed, together with the H$\alpha$ images of the
flaring area delimited by the appropriate diaphragm.
The spectral flux data are
flat-fielded, calibrated and processed to be compared with model predictions.
Our analysis of the data proves that the described device is sufficiently sensitive to
detect variations in the Balmer continuum during solar flares. Assuming that the Balmer-continuum
kernels have at least a similar size as those visible in H$\alpha$, we find
the flux increase in the Balmer continuum to reach 230\% \-- 550
\% of the quiet continuum during the observed X-class flare.
We also found temporal changes in the Balmer continuum flux starting well
before the onset of the flare in H$\alpha$.
\end{abstract}


   \end{opening}
%
%
%
\section{Introduction}

Spectral observations of solar flares have been carried out in
lines of various species, but also in the continua ranging from
EUV up to microwaves. Several recent studies indicate that the
continuum radiation represents a significant portion  of the total
energy which is deposited in the lower atmospheric layers \textit{e.g.} by
the electron beams (\citealp{Watanabe13}; \citealp*{Kerr14};
\citealp{Milligan14}). Among these continua, the most intriguing is
the optical continuum detected between the hydrogen Balmer-jump
region and near infrared spectral bands. Flares exhibiting such
emissions are called 'white-light flares' (WLF). \citet{Neidig89}
defined solar WLF as components of flares that are visible in the
optical continuum or integrated light. The WLF emission appears as
patches, waves, or ribbons often containing bright kernels smaller
than 3 arcsec. \citet*{Nei+Cli83} also stated that WLFs are
associated with more energetic EUV and X-ray flares.
\citet{Jess+all2008} measured 300\% increase in solar white light
continua lasting 2 min and being co-temporal and co-spatial with
the flare event. They suggested that the creation of white-light
emission is a common feature of all solar flares including
less energetic ones. This is in contrast to \citet{Fletcher+07} and
\citet*{Fletcher+Hud07} who claimed that the mechanisms of small WLFs  are
still open to debate.

However, the detection of WLFs is difficult because they
are short-lived events usually co-temporal with the
impulsive onset of flares and not well correlated with \textit{e.g.} the
H$\alpha$ line kernels. Moreover, there are apparently two types
of WLFs which have different spectral signatures (\citealp{Ding2007}):
(I) optical continuum enhancement over a broad wavelength range which is assigned to
a photospheric temperature increase below the temperature-minimum
region (photospheric H$^-$ continuum), and (II) hydrogen
recombination continua produced mainly in the flaring
chromosphere. The latter continua span a wide range of
wavelengths, from EUV (Lyman continuum) up to infrared. Having
good spectral coverage by ground-based or space instruments, one
could in principle disentangle between these two types of WLFs.
But this is not usually the case in optical flare observations
where typically one or a few wavelength channels are used, like three
narrow-band channels on the {\it Solar Optical Telescope} (SOT) of {\it Hinode} (\citealp*{Kerr14}). While the
broad-band continuum is frequently fitted by a black-body curve in
order to estimate its temperature (see the above references), such
a fit leads to too low enhancement of the Balmer continuum which
contradicts the numerical simulations (\citealp{Kleint15}). Various
attempts have been made to detect the Balmer continuum enhancement
during flares, but the results are rather controversial. In some
case the Balmer jump was detected (\textit{e.g.} \citealp{Zirin81};
\citealp{Hiei82}), in others only a smooth transition from the
so-called 'blue continuum' (\citealp*{Donati-Falchi85};
\citealp{Kowal15}) to the Balmer continuum was detected (see also
summary by \citealp{Neidig89}). Many flare spectra with good spectral
resolution have been collected at the Ond\v{r}ejov Observatory
during the sixties, but \citet{Svestka66} claimed that there was no
evidence of the Balmer-continuum enhancement. However, this might be
because of the photographic technique used at that time. Only
recently, \citet*{Hei+Kle2014} have found a clear signature of the
continuum enhancement in the far wing of Mg {\sc ii} h-line during an
X1-class flare observed by the {\it Interface Region Imaging Spectrograph} (IRIS) satellite and they
attributed this emission to the Balmer continuum. This is
promising, but in order to determine reliably the spectral shape
of the continuum, we need simultaneous observations of the Balmer
continuum in more than one narrow-band channel. We therefore
started a systematic observing program of detecting the Balmer continuum
from ground and we designed a novel instrument as
described in this paper.

\section{Instrument}

As mentioned above, WLFs are usually short-lived brightenings and
thus it is very difficult to detect them at the right position (with
the spectrograph slit) and at the right time. It is also difficult
to detect them in the integrated solar flux by observing the Sun
as a star - the contrast is too low because of the bright optical
continuum of the whole solar disk (contrary to EUV continua
observed by the {\it Extreme Ultraviolet Variability Experiment} (EVE) on the {\it Solar Dynamics Observatory} (SDO); see \citet{Milligan14} and references
therein.) Note that on cool dMe stars the situation is much more
favorable; see \citet{Kowal2013}. We therefore suggest to use a
special instrumental setup which allows us to detect the optical
spectra in certain wavelength range only from a limited area of
the solar disk covered by an active region where the flare
occurrence is expected. We have considered the following
requirements for the ground-based feeding telescope and for the
new post focal spectroscopic instrument:
\begin{itemize}
 \item high ratio \textit{D/f} of the objective to have enough light ($D$ and $f$ are the
diameter and focal length, respectively)
 \item stable guiding system to follow selected active region for a long
 time
 \item continuous imaging of the active region through the H$\alpha$
 filter
 \item imaging of the diaphragm position on the  H$\alpha$
 filtergram
 \item broad-band spectrometer with high sensitivity in the continuum
 \item high cadence up to 50 recordings per second
 \item an accurate image selector that picks up a proper target region on the solar disk
 \end{itemize}
 Note that the seeing issues, extremely important for slit instruments, are not critical here because
 of  the flux
integration of the whole active-region/flare area.
 During the device development period we found that all the
 above mentioned parameters of the experiment are
 important, but the most critical is the way how the area to be measured is selected  in the optical
 system.

 As the ground-based telescope we used the large horizontal
 telescope \textit{Horizontal Sonnen Forschungs Anlage 2} (HSFA2) of the Ond\v{r}ejov Observatory described by \citet{Kotrc09} with a Jensch\--type coelostat
 located 6 m above the ground
 and with the main objective mirror $D/f$=50/3500, where both $D$ and $f$ are in cm. The diameter of the
solar disk at the focal plane of the HSFA2 telescope is 32 cm.
 A new device called the Image Selector was placed into the focal plane of the telescope (see
 Figure~\ref{schema}) to feed both the H$\alpha$ filter
 and the spectrometer HR4000 (see below).
  \begin{figure}[h]
   \centering
   \includegraphics[width=0.90\textwidth,clip=]{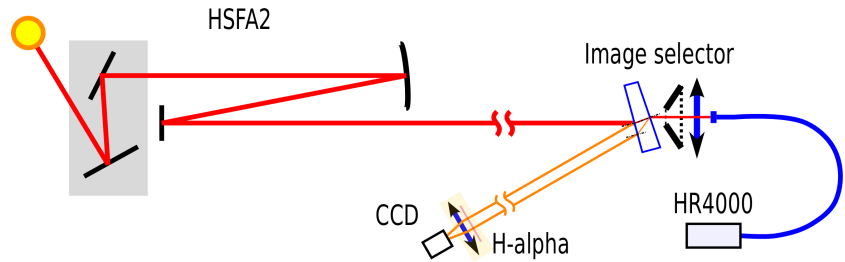}
   \caption{ Optical scheme of the Image Selector located
   in the focal
   plane of the HSFA2 telescope. The selector defines a circular area around the
active region or flare ribbons from which the light is integrated and sent by the fiber
optics to the spectrometer HR4000.
  }
              \label{schema}
    \end{figure}

The light from the telescope HSFA2 enters the Image Selector
through an inclined glass wedge, which is covered with coatings
to reduce the light above 400 nm where the signal becomes much
stronger compared to the Balmer limit region. Behind the
wedge, there is a rotating wheel with seven circular diaphragms each
having a central entrance hole and an outer rim which is highly
polished to ensure the back-reflection of unused light. The internal shell of each
diaphragm has a conical shape to reduce the parasitic light. The
diaphragms have the holes 4, 8, 10, 12, 14, 16, and 18
mm in diameter which correspond to 23, 46, 57, 69, 80, 92, and 103 arcsec of
the focal image, respectively. Only light from the portion of the
solar image which is projected onto the diaphragm hole enters the
flux detection optics. This is achieved by using a collecting lens
(vertical line with two arrows in Figure 1) and an optical fiber
behind it which feeds the spectrometer HR4000. The grating and the
detector of the spectrometer were selected to have up to 0.02 nm
(FWHM) spectral resolution covering 350 \--- 440 nm spectral
range. The spectrometer has a linear CCD detector with 3648 pixels.
Depending on the amount of light, the spectrometer can achieve a
cadence of up to 50 frames per second. The spectrometer HR4000 is a
low-cost low-dispersion compact instrument made for laboratory and
educational purposes by Ocean Optics Co., USA (the instrument
comes with the software and can be easily connected to a PC). It
detects very well the solar continuum radiation and also resolves
the strongest Fraunhofer lines which typically go into emission
during flares.

Part of the light is reflected from
the first optical surface (the front side of the wedge) of the Image
Selector and together with the enhanced reflection from the
circular rim of the diaphragm, the beam enters the
H$\alpha$ telescope - this provides information about the flaring
activity within the monitored region. The inner circle of the
bright circular rim projected onto the H$\alpha$ images
delimits the region where the spectral flux is measured. The anti-reflection coatings
both on the glass-wedge surfaces and on the
collecting lens feeding the optical fiber reduce the scattered
light to minimum. The usual cadence of the H$\alpha$ monitoring
is about one image per second. The rotation of the diaphragm
wheel, the spectrometer itself, and the H$\alpha$ telescope are controlled by
a PC which also stores all the data. Spectra are corrected for dark
frames and are flat\--fielded.

\section{Measurements}
We used the following scheme of observations. According to a solar
flare prediction and actual activity we have selected an active region
and projected it to the center of a diaphragm of an appropriate
size. Before that, dark frames and flat-field images were taken.
Then the selected active region was tracked and the exposure times
both at the spectrometer detector and the H$\alpha$ camera were
set up. As the amount of light changes considerably with the
altitude of the Sun, the flat field images and the exposure times
have to be adjusted for new data series.

Shortly after the active region NOAA 12087 appeared above the
south-east limb, three X-class flares were detected by the {\it
Geostationary Operational Environmental Satellites} (GOES). None
of them was accompanied by CME. The first two, X2.2
(SOL2014\--06\--10T11:42) and X1.5 (SOL2014\--06\--10T12:52)
flares, were observed with our instrument on 10 June 2014, but not
from the beginning of the flare. In addition, their positions
close to the eastern limb (S19E81 and S20E89) complicated the
integrated flux measurements due to the emission lines from the
nearby limb region. On 11 June 2014 the AR 12087 was at S18E57
(-755 arcsec, -297 arcsec; $\mu$=0.39) and was of $\beta\delta$
Hale class, with five spots and produced an X1.0 flare
(SOL2014\--06\--11T09:06). This flare was observed during all its
duration, in total 25 min 25 s, including several minutes of
pre-flare phase.

\begin{figure}[ht]
   \centering
   \includegraphics[width=0.92\textwidth,clip=]{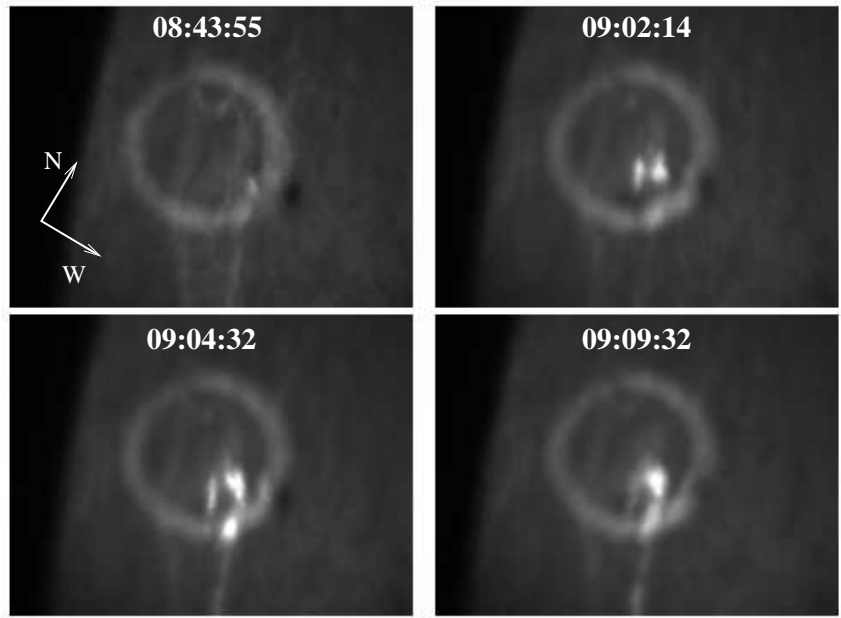}
   \caption{ H$\alpha$ images of the flaring region with flare kernels.
   The circular rims are reflections from the front part of the diaphragm delimiting the entrance window to the spectrometer.
   Faint vertical structures correspond to filamentary
   brightenings occurring in the active region.
  }
              \label{slit-jaw}
    \end{figure}

In Figure~\ref{slit-jaw} four H$\alpha$ filtergrams
from reflection on the Image Selector during the
flare of 11 June (SOL2014\--06\--11T09:06)
are shown. Note the bright circular
reflection from the polished front of the diaphragm of diameter 10
mm (57 arcsec) which was selected for this active region. The
south-eastern solar limb was positioned to the left of the flare position,
S18E57. On the first image (top left) no flare in H$\alpha$ was
yet detected.

A small portion of light from the solar image is reflected from the
front surface of the Image Selector, and together with the circular
reflection on the front part of the diaphragm it enters the
H$\alpha$ camera. During the flare evolution we observed the
developing bright kernels in H$\alpha$ with a
cadence 1 secs. The spectrometer cadence was 11 spectra per second
while its integration time was 30 ms.

One of the three flare kernels (the most south\--western one)
was projected on the front part of the diaphragm and thus it had
no contribution to the spectral flux measurement. Without moving the
image of the solar disk during the flare observation we kept the
background radiation unchanged. This is particularly
important when the flare is close
to the solar limb.

\begin{figure}[h]
   \centering
   \includegraphics[width=1\textwidth,clip=]{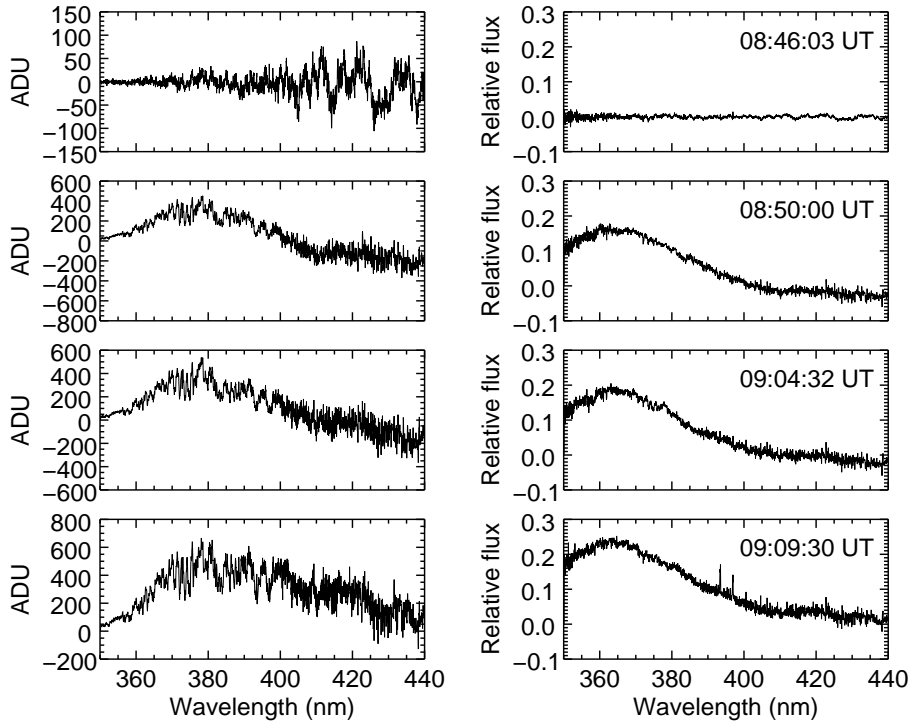}
   \caption{The net and relative solar flux excesses at four instances of the measurement.
   The times are close to those of four H$\alpha$ images displayed in
   Figure 2. In the left column the spectral flux of
   the quiet region $Q$ is subtracted from the flare flux $F$ measured during the
   flare $(F\--Q)$. In the right column the result of subtraction is divided
   by the quiet flux, $(F\--Q)/Q = F/Q - 1$.
  }
              \label{excesy-all}
    \end{figure}

In Figure~\ref{excesy-all} we can see examples of the net increase in
the spectral flux as collected by the Image Selector and detected by the
spectrometer. The spectral excess was prominent during the flare
from 350 nm up to 410 nm. In the upper pair of images the
measurement before the flare onset can be seen. Note the Ca~{\sc ii}
lines in absorption. But later, around the flare maximum,
both the calcium as well as higher Balmer-series lines
were visible in emission. We thus see that selected Fraunhofer lines
changed their shapes gradually from absorbtion to
emission (see the right column in Figure~\ref{excesy-all}).

The maximum spectral increase occurs at the wavelength 364.6~nm
which is the Balmer limit (Balmer jump) and the wavelengths
shorter than that is known as the Balmer continuum. The
spectral flux excess during the flare was detected in all the measured
range of wavelengths below 410 nm, which is often called a "blue
continuum" (\citealp*{Donati-Falchi85}). Both the net and the
relative excess can be seen at the lower three pairs of spectra in
Figure~\ref{excesy-all}.

\begin{figure}[h]
   \centering
   \includegraphics[width=0.75\textwidth,clip=true, trim=0 0 5cm 0]{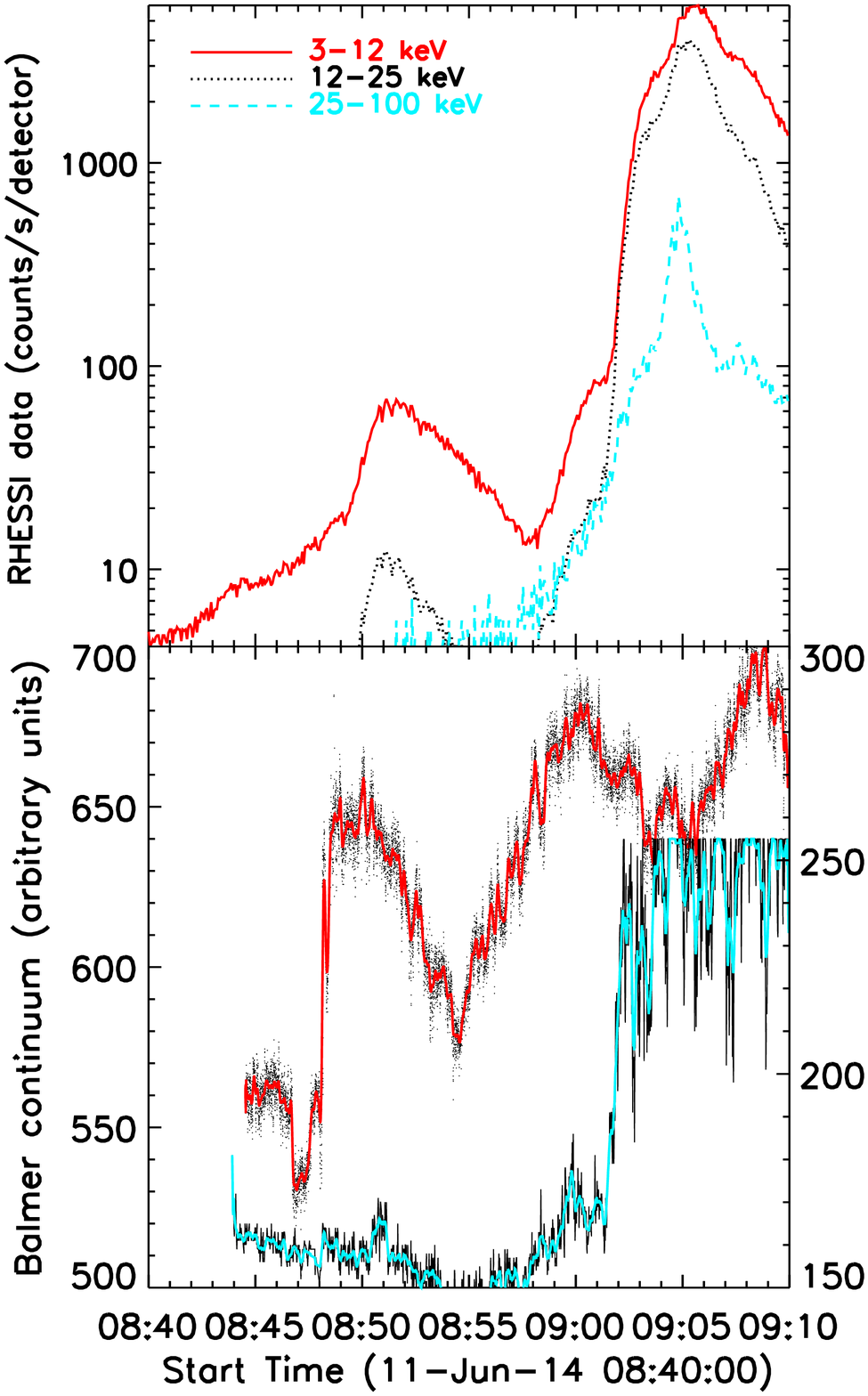}
   \caption{Light curves of the H$\alpha$ signal proxy (blue) and the Balmer
   continuum flux (red) in the bottom panel and the X-ray flux in three channels from RHESSI (top panel).
   The H$\alpha$ light curve shows the maximum value detected
   over all the image area including the area out of the diaphragm.
  }
              \label{composed}
    \end{figure}

We studied also the light curve of the Balmer-continuum channel
and compared it with the light curve of the H$\alpha$ signal. The
latter was defined as the maximum signal in ADU (analog-to-digital
converter unit) in the H$\alpha$ camera. This value is taken as a
proxy of the H$\alpha$ signal. It is reliable before it reaches
the maximum value equivalent to the 8 bit camera, then the signal
becomes saturated. Both these light curves can be seen in the
lower part of Figure~\ref{composed}.  Unfortunately for this flare
we found no other observation of H$\alpha$ light curve with a
comparable time resolution.

\section{Analysis}
In each processed H$\alpha$ image we marked the circle delimiting
the measured area. Inside the circle we specify flare kernels as
an area where we measured the signal equal to at least the double
median value in the whole image.  Then the ratio of the number of
pixels in the flare kernels to the total number of pixels inside
the measured area gives the ratio between the flare area and that
covered by the Image Selector circular diaphragm. For the analyzed
flare this was in the interval 3.0 \-- 5.8 \%. The measured
relative excess of the continuum flux at 350 nm was between 13 \--
17 \%,  while at the maxima of the curve it was  between 17 \-- 21
\%
 (see Figure~\ref{excesy-all}). Assuming that the flare kernels
emitting in the Balmer continuum were of the same size as those in
H$\alpha$, the reconstructed relative excess of the Balmer
continuum should be in the range from 230 to 550~\% and this will
be even larger for smaller kernels.

To compare the measured spectral flux enhancement with the models of
Heinzel and Ka\v{s}parov\'{a} (2015) we calibrated our measurements using the
continuum intensities tabulated by \citet{Allen76}. The intensity of
the quiet Sun is related to the signal measured before the first
signatures of a flare. After the flaring kernels have appeared on
H$\alpha$ images, we subtracted the spectral signal in
the Balmer-continuum channel for the quiet Sun from the signal in the
same channel at the flare time. Finally we divided this value with
the ratio between the flaring area and the whole measured circular
area as calculated from the respective H$\alpha$ image. After taking into account the center-to-limb variations,
we obtained the intensity of
the flare kernels.
The recalculated specific intensity of the Balmer continuum at 350 nm in this flare was from 9 to 15 (or from 14 to 23)$\times
10^{-6}$ erg s$^{-1}$ cm$^{-2}$ sr$^{-1}$ Hz$^{-1}$ according to
selected type of calibration. These values are consistent with the
intensity enhancement at 350 nm, computed by Heinzel and Ka\v{s}parov\'{a} (2015) for
the static flare models of \citet*{RC83} with a flux of 10$^{11}$ erg
s$^{-1}$ cm$^{-2}$, spectral index 3-5, and coronal pressure
10$^2$-10$^3$ dyn cm$^{-2}$. Such models are representative of a
strong flare with the electron beam precipitating to chromospheric
layers. However, in this exploratory study we did not intend to
analyze our new measurements around the Balmer limit at 364.6~nm
for the following reasons. It is well known that during flares,
highest members of the Balmer series merge together due to
significant Stark broadening and a quasi-continuum is formed in
front of the Balmer limit as a result of the lowering of the
hydrogen ionization potential (see \citet*{Donati-Falchi85} and
references therein). Detailed discussion on the formation of the
Balmer continuum around the limit can be found in
\citet{Kowal15}. Based on our measurements, we interpret the flare
continuum enhancement at 364 nm as the emission of optically-thin
Balmer recombination continuum not affected by rather complex
processes which take place around the Balmer-continuum limit. Our
result is consistent with the analysis of IRIS spectra taken
during an X-class flare in the near-UV wavelengths - see Heinzel and Ka\v{s}parov\'{a} (2015).
The models which fit best our observations have the optical
thickness at the Balmer jump lower than unity at $\mu$=0.39
(position of our flare), typically between 0.1 - 0.7. This is the
optical thickness of the chromospheric layer in which the
Balmer-continuum enhancement is produced. However, the total
optical thickness of the Balmer continuum including the whole
photosphere is much larger than unity. In the case of stellar flares
(dMe stars) it is likely that the chromospheric Balmer continuum is
optically thick in models with strong fluxes, stronger than
considered in our modeling (see \textit{e.g.} Kowalski {\it et al.}, 2015).

 The light curves
of the Balmer continuum and of the H$\alpha$ proxy were observed
since 08:43 UT, \textit{i.e.} 16 min before the H$\alpha$ flare steep rise
till 9:10 UT. In Figure~\ref{composed} we can compare them
with the X\--ray light curves as measured by the {\it Reuven Ramaty High-Energy Solar Spectroscopic Imager} (RHESSI; \citealp{Lin:2002aa}). During
the event all the light curves varied substantially, but,
there is no precise correlation between them. The Balmer-continuum
emission slightly decreased at 8:47 UT (8 min before the flare
started in H$\alpha$). A similar decrease was also detected in the
Lyman continuum with the SDO/EVE spectrometer which
detects the total flux from the whole flare region (R.Milligan,
private communication).
Then the Balmer-continuum sharply increased from 530 to 650 ADU peaking at
8:50 UT. This local maximum of the Balmer continuum flux coincided
with a gradual rise of the 3 \-- 12 keV channel of RHESSI signal having its
maximum at 8:52 UT. Thus the first Balmer continuum changes occured about
10 \-- 12 min before the flare in H$\alpha$ began.
We can see a rough coincidence among the 12-25 keV signal, H$\alpha$ intensity,
and the Balmer continuum peaks around 08:51 UT. The 25-100 keV peak
coincided with the double peak in Balmer continuum flux. The
differences can be due to different integration areas of RHESSI
and our Image Selector on the solar disk containing the flare
kernels. In Figure 2 we see that one bright H$\alpha$ kernel appeared
around the flare maximum just on the bright rim of the Image
Selector and this represents a missing signal in the
Balmer-continuum flux around this time where we get a decrease
between the two peaks in Figure 4. The Balmer signal at the RHESSI hard X-ray maximum is about 650 units.
This corresponds to two kernels within the diaphragm, so that a mean signal
from one kernel is more than 300 units. Adding this value as a missing signal from
the third kernel will increase the total signal to more than 900 units and thus will produce
a clear peak of the Balmer signal at time of the hard X-ray maximum. Note that \citet*{Hei+Kle2014}
found a good correlation between Balmer-continuum enhancement and
RHESSI signal.

\section{Summary and Conclusions}
We have developed a new post-focal instrument called the Image
Selector, attached to a broad-band spectrograph HR4000 for
measurements of the spectral flux during solar flares. The basic
idea behind this experiment is to integrate the continuum flux
from the whole flare area. This eliminates problems with detecting
a WLF kernel on the spectrograph slit and also avoids difficulties
caused by the seeing. Our setup is similar to that for observing
the Sun as a star, but by using only a small area around the flare
instead of the whole solar disk, we substantially increase the
continuum enhancement contrast detected during flares. Moreover,
we are recording the broad-band continuum spectra which was done
recently only in case of flare stars. The Image Selector technique
has proven to be sensitive enough to detect solar flux increase in
the spectral region of the hydrogen Balmer continuum. Relative
excess of the Balmer continuum flux (irradiance) in the measured
X1.0 flare reached at least 230 \% \-- 550 \% as compared to the
pre-flare situation. The reconstructed Balmer continuum specific
intensity (radiance) at 350 nm for this flare was from 9 to 15 (or
from 14 to 23)$\times 10^{-6}$ erg s$^{-1}$ cm$^{-2}$ sr$^{-1}$
Hz$^{-1}$, according to the selected type of calibration. This is
in good agreement with theoretical synthetic spectra computed by
Heinzel and Ka\v{s}parov\'{a} (2015, in preparation) who predicted
Balmer-continuum enhancement in a strong flare heated by the
electron beam precipitating into lower atmospheric layers and is
also consistent with recent detections by IRIS
(\citealp*{Hei+Kle2014}). We have also shown the temporal
variations of the Balmer-continuum flux and compared them with the
H$\alpha$ proxy light curve and with RHESSI hard X-ray fluxes, but
more observations are certainly needed to estimate the precise
degree of correlations. We found temporal changes in Balmer
continuum flux starting even 16 min before the onset of the flare
in H$\alpha$. If it is found out to be true, this information
could be used for a short time prediction of the flare activity in
a given active region.   During the impulsive phase of the flare
(\textit{i.e.} after the flare onset), the hydrogen continua are
formed by radiative recombination while the H$\alpha$ line
formation is mostly due to collisional (thermal and non-thermal)
excitation. Before the flare onset in H$\alpha$, the situation can
be more complex and this will require new simulations.

 Having an instrument with high cadence measurements
of the spectral flux in blue continua emitted from a limited
flaring region, we would be able to study individual spectral
regions responsible for different mechanisms of the creation of
the continua. Then, using a statistically larger set of
observations we could sort out individual WLFs into group I or II.

As our first observational results are very promising, we plan to
install behind the Image Selector another similar Ocean Optics
spectrometer to cover longer wavelengths. This will allow us to
detect the hydrogen Paschen continuum including the Paschen limit,
as well as the whole optical continuum which can be enhanced due
to a photospheric temperature increase (H$^-$ continuum). Both
such low-cost low-dispersion spectrometers will operate
simultaneously, together with the H$\alpha$ imaging. Because of
its compactness and low weight, it will be possible to mount this
system on other solar telescopes. Such relatively simple
observations will help to clarify the role of the continuum
emission during solar flares and to prove the statement of
\citet{Jess+all2008} that the WLF emission is a common feature
of all solar flares including less energetic ones.  It will be
promising to use the {\it Optical and Near-infrared Solar Eruption
Tracer} (ONSET) telescope (Fang {\it et al.}, 2013). A very
interesting observations mentioned by  \citet{Cheng2015} and
\citet{Hao2012} show that the 3600 \AA\, flare kernel is much
smaller than the
 H$\alpha$ one. If our analysis is applied, it will lead to a strong Balmer-continuum intensity enhancement as derived from the flux, consistent
with our discussion in Section 4. Future joint observations with
our instrument and with ONSET would be very valuable.
 Further observations of the optical flare
continua should bring a new insight into the role of accelerated
particles, ionization and recombination processes, the flare
heating, and its overall energetics.

\begin{acknowledgements}
The research leading to these results has received funding from
the EC Program
FP7/2007-2013 under the F-CHROMA grant agreement No. 606862,
and funding from the People Program - Marie Currie Actions of
FP7/2007-2013 under the REA
grant agreement No. 295272 (Radiosun). Grant from the Czech
Funding Agency (GACR) No. P209/12/1652 also partially supported this project.

\end{acknowledgements}

\bibliography{kotrc-v2}
\bibliographystyle{spr-mp-sola}
\end{article}
\end{document}